# Symmetry-enforced planar nodal chain phonons in non-symmorphic materials

**Hong-Ao Yang, Hao-Yu Wei, and Bing-Yang Cao***

Key Laboratory for Thermal Science and Power Engineering of Ministry of Education, Department of Engineering Mechanics, Tsinghua University, Beijing 100084, China

*Corresponding author: Tel/Fax: +86-10-6279-4531; E-mail: caoby@tsinghua.edu.cn

# Abstract

Topological semimetal states which are constrained by symmetries and give birth to innovative excitations are the frontiers of topological quantum matter. Nodal chains in which two nodal rings connect at one point were first discovered in non-symmorphic electronic systems and then generalized to symmorphic phononic systems. In this work, we identify a new class of planar nodal chains in non-symmorphic phononic systems, where the connecting rings lie in the same plane. The constituting nodal rings are protected by mirror symmetry, their intersection is guaranteed by the combination of time-reversal and non-symmorphic two-fold screw symmetry. In addition, the connecting points are four-fold degenerate while those in previous works are two-fold degenerate. We searched all 230 space groups and found 8 space groups that can host the proposed planar nodal chain phonons. Taking wurtzite GaN (space group No.186) as an example, the planar nodal chain is confirmed by first-principles calculations. The planar nodal chains result in two distinct classes of drumhead surface. The first category lies on the [10(-1)0] surface Brillouin zone and the second lies on the [0001] surface Brillouin zone. Our finding reveals a class of planar nodal chains in non-symmorphic phononic systems, expands the catalog of topological nodal chains, and enriches the family of topological surface states.

# 1. Introduction

Topological quantum states attract much attention in condensed matter physics nowadays since they delineate quantum phase transition upon change of topological invariants[1-4]. In the past decade, topological semimetal states, in which the low energy excitations near band degenerate points can be described by relativistic Dirac and Weyl equations, were found in 3D solids[5-8]. Weyl points can be viewed as sources and sinks of Berry flux from the aspect of topology, resulting in Fermi arc states in the surface Brillouin zone[9]. Dirac points formed by four-dimensional irreducible representations and accidental band crossing were found[10-12]. Moreover, higher dimensional degeneracies like nodal rings[13-15], nodal chains[16], nodal knots[17], and nodal surfaces[18,19] are classified based on symmetry analysis. The symmetry constraints of lattices are essential for band degeneracies of dimensions higher than zero. A two-fold band degenerate point can exist without lattice symmetry, while a two-fold band degenerate line preserves under the protection of mirror symmetry in the simplest case. Drumhead-like flat surface states are found in nodal ring semimetals[20], which may offer a route to superconductivity. However, the detection of topological semimetal states is restricted because of the constrain of the Fermi level and the symmetry-breaking due to spin-orbit coupling.

Phonons, as another elementary excitation in solids, can also display nontrivial degeneracy and topological effects. They play an important role in heat transport, electron-phonon energy transfer, and superconductivity. Phonons are bosons, so the entire frequency range of phonons can be excited and observed in experiments. The analog of semimetal states in phonons is not restricted by the Fermi level, significantly enriching potential topological phononic materials. By introducing pseudospin, quantum spin Hall like states and helical edge states have been established[21] and verified in acoustic metamaterials[22-25]. Due to the absence of the spin degree of freedom and Kramers degeneracy, low energy excitations near band crossing points are mostly Weyl semimetals. According to the dimension of degeneracy, they can be divided into 0-D Weyl points[26-28], 1-D Weyl nodal lines[29-32], and 2-D Weyl nodal surfaces[33-35]. Topologically protected surface states have been confirmed in experiments[29,36]. Recently, phononic nodal lines of different shapes have been reported, including helical nodal lines[29], straight nodal lines[30,31], and nodal rings[32,37–39].

Taking more symmetry constraints into consideration, nodal rings are found to build up chain-like geometries. Nodal chains formed by intersecting nodal rings from different planes are found in electronic systems[16], where the intersection is guaranteed by non-symmorphic symmetries and is immune to spin-orbit coupling. The concept of nodal chains is soon after introduced to phononic systems[37,40], whereas the intersection is protected by symmorphic symmetries. However, it remains unclear whether non-symmorphic symmetry-protected nodal chains can exist in phononic systems, which is critical given the fact that non-symmorphic space groups count up to 157 out of 230 space groups. In addition, currently known nodal chains consist of nodal rings from different planes, while nodal chains composed of nodal rings lying in one plane are still missing.

In this work, we identify a new class of planar nodal chains protected by non-symmorphic symmetries in phononic systems. The constituting nodal rings lie in the same mirror-invariant plane, thus named planar nodal chains. Their intersection point is a four-fold degenerate point that is protected from gapping by a combination of time-reversal symmetry and non-symmorphic screw symmetry. The symmetry conditions do not apply to electronic systems due to the spin degree of freedom. We identified 8 space groups that can host the planar nodal chain phonons, including known realistic materials. Based on first-principles calculations, we demonstrated the existence of planar nodal chain phonons in a representative material, wurtzite GaN. In addition, the planar nodal chain leads to arc states both on surface Brillouin zones oblique to the mirror plane and perpendicular to the mirror plane, facilitating experimental verification.

## 2. Symmetry Constraints

Nodal rings are one-dimensional degeneracies in three-dimensional momentum space. The codimension is two, so extra crystal symmetries are needed to stabilize the nodal lines (rings). One of them is mirror symmetry $\mathcal{M}_x: (x, y, z, t) \mapsto (-x, y, z, t)$. Here we take $x = 0$ as the mirror invariant plane. In phonon systems without the spin degree of freedom, $\mathcal{M}_x^2 = 1$, and the eigenvalues of $\mathcal{M}_x$ are $\pm 1$. Eigenstates with opposite mirror eigenvalues are forbidden to hybridize, enabling two-fold band degeneracy, that is, the nodal ring, as shown in Fig.1(a). We refer to the degenerate states as $|\phi_+\rangle$ and $|\phi_-\rangle$, where the subscripts denote mirror eigenvalues. Introducing more symmetry, the band degeneracy can expand to four-fold at certain points, as explained below.

A combination of time-reversal symmetry $\mathcal{T}$ and two-fold screw symmetry $\mathcal{S}_z$ guarantees Kramers degeneracy in the $k_z = \pi$ plane in spinless systems, forming nodal surfaces[18]. We demonstrate as follows that $\mathcal{T}$, $\mathcal{S}_z$, and $\mathcal{M}_x$ together generate two sets of two-fold irreducible representations in the $(k_x = 0, k_z = \pi)$ high symmetry line, and further lead to a four-fold degenerate point. In real space, $\mathcal{T}$ and $\mathcal{S}_z$ act as

$$\begin{aligned}\mathcal{T}&: (x, y, z, t) \mapsto (x, y, z, -t) \\ \mathcal{S}_z&: (x, y, z, t) \mapsto \left(-x, -y, z + \frac{1}{2}, t\right).\end{aligned} \quad (1)$$

While in momentum space,

$$\begin{aligned}\mathcal{T}&: (k_x, k_y, k_z) \mapsto (-k_x, -k_y, -k_z) \\ \mathcal{S}_z&: (k_x, k_y, k_z) \mapsto (-k_x, -k_y, k_z).\end{aligned} \quad (2)$$

We have $[\mathcal{T}, \mathcal{S}_z] = 0$, and the combined symmetry $\mathcal{S} = \mathcal{T}\mathcal{S}_z$ acts as

$$\begin{aligned}\mathcal{S}&: (x, y, z, t) \mapsto \left(-x, -y, z + \frac{1}{2}, -t\right) \\ (k_x, k_y, k_z) &\mapsto (k_x, k_y, -k_z).\end{aligned} \quad (3)$$

Notice that $\mathcal{S}_z^2$ is equal to a translation along the $z$ axis by a lattice constant, so we have $\mathcal{S}^2 =$

$S_z^2 = -1$ when $k_z = \pi$. Meanwhile, $S$ preserves $k$ vector since $k_z = \pi$ is equivalent to $k_z = -\pi$. So $S$ ensures Kramers degeneracy in the $k_z = \pi$ plane, making all representations two-fold irreducible. Adding this condition to the mirror plane where $k_x = 0$, $|\phi_+\rangle$ and $|\phi_-\rangle$ are accompanied by $S|\phi_+\rangle$ and $S|\phi_-\rangle$, respectively. The following four states are forbidden from hybridization along the high symmetry line ($k_x = 0, k_z = \pi$),

$$|\phi_+\rangle, |\phi_-\rangle, S|\phi_+\rangle, S|\phi_-\rangle.$$

To prove their orthogonality, we take $|\phi_+\rangle$ as an example. $|\phi_+\rangle$ and $S|\phi_+\rangle$ are Kramers pairs, so they are immune from hybridization. $|\phi_+\rangle$ and $|\phi_-\rangle$ are protected by opposite mirror eigenvalue. Since $S$ and $\mathcal{M}_x$ commute, $S|\phi_-\rangle$ has the mirror eigenvalue -1, preventing it from hybridizing with $|\phi_+\rangle$. Therefore, the four states are capable to produce a four-fold degenerate point in the ($k_x = 0, k_z = \pi$) high symmetry line, see Fig.1(b).

The accidental band crossing mechanism guarantees that the four-fold degenerate point is stable in a wide range of the parameter space[11], as long as the symmetries $\mathcal{M}_x$, $\mathcal{T}$, and $S_z$ preserve. Unlike Dirac points which are isolated gap-closing points, the four-fold degenerate point serves as the crossing point of multiple nodal rings. Specifically, three kinds of nodal rings each between two adjacent bands are presented in wurtzite GaN, which will be verified by calculation in the results section, as shown in Fig.1(c). All intersecting nodal rings are in the mirror plane $k_x = 0$, thus called the planar nodal chain.

The discussion above can be generalized to space groups in which the three symmetries $\mathcal{M}_\alpha$, $S_\beta$, and $\mathcal{T}$ are present, where $\alpha$ and $\beta$ are two nonparallel directions, denoting the mirror plane direction and the screw axis, respectively. Since the screw symmetry is non-symmorphic, the symmetry conditions only apply to non-symmorphic space groups. In addition, the mirror symmetry-protected nodal rings break down in the presence of spin-orbit coupling, so the symmetry conditions are not valid in electronic systems. Among all 230 space groups, we identified 8 candidates for the planar nodal chain phonons, as listed in Table.1. Some realistic materials are also given, of which Si$_2$N$_2$O is a kind of ceramics, GaN and AlN are wide band-gap semiconductors, and NaS is used in sodium–sulfur batteries. In the following, we will take wurtzite GaN (space group No.186) as an example to illustrate the planar nodal chain phonons. Wurtzite GaN is widely used in high-power electronics due to its wide band gap and high carrier density. Yet their performance and lifetime are restricted by heat dissipation efficiency. The exploration of topological phonon states offers new opportunities for tuning phonon transport in GaN devices.

## 3. Computation Details

Wurtzite GaN is studied using density functional theory (DFT). All DFT calculations are performed with the Vienna Ab initio Simulation Package (VASP)[41]. Projective augmented wave pseudopotential[42] in Perdew–Burke–Ernzerhof[43] formalism is employed for the exchange-correlation functional. Lattice constants of wurtzite GaN are optimized under constraints of symmetry first. After relaxation, density-functional perturbation theory is performed on a $4 \times 4 \times 3$ supercell with a Γ-centered $4 \times 4 \times 3$ $k$ mesh to extract the second-order force constants. The Phonopy package[44] is used to construct phonon dynamical matrices. Iterative Green's function method[45] as implemented in WannierTools[46] is adopted to obtain the phonon surface state spectrum.

## 4. Results and Discussion

Wurtzite GaN belongs to space group No.186 ($P6_3mc$). Its unit cell is shown in Fig.2(a). There are four atoms in the unit cell, thus are twelve phonon branches. The optimized lattice constants are $a = 3.216$Å and $c = 5.240$Å. Fig.2(b) shows the first Brillouin zone of bulk GaN and the projected surface Brillouin zones on the [0001] surface and the [10(-1)0] surface. High symmetry points in bulk and surface Brillouin zones are labeled. The generators of space group No.186 are mirror symmetry $\mathcal{M}_x$, two-fold screw symmetry $\mathcal{S}_{2z}$, and three-fold rotation symmetry $\mathcal{C}_{3z}$. As discussed above, the mirror symmetry $\mathcal{M}_x$ is capable to protect the nodal rings in the mirror invariant plane Γ-A-L-M. The two-fold screw symmetry $\mathcal{S}_{2z}$ and time-reversal symmetry $\mathcal{T}$ ensure Kramers degeneracy in the $k_z = \pi$ plane, i.e., the A-L-H plane, making it a nodal surface for all bands, which means every phonon band in the $k_z = \pi$ plane is two-fold degenerate. $\mathcal{M}_x$, $\mathcal{S}_{2z}$, and $\mathcal{T}$ together generate a planar nodal chain in the Γ-A-L-M plane. Fig.2(c) displays the phonon dispersion of wurtzite GaN along high symmetry paths, along with the nodal plane A-L-H and the four-fold degenerate point D. Calculation results show the coordinate of the D point is (0.384, 0, 0.5) in reciprocal lattice units. The degenerate frequency at the D point is 4.93 THz.

To verify the existence of the nodal rings and the planar nodal chains, we thoroughly calculate the phonon dispersion in three dimensions in $k$ space. The exact shape of the planar nodal chain is plotted in Fig.3(a), which is similar to that in Fig.1(c), that is, three kinds of nodal rings intersect at D point. Although not shown, there are also planar nodal chains in five other directions due to six-fold screw symmetry, which is a combination of two-fold screw symmetry and three-fold rotational symmetry. The blue lines are the degenerate lines of the first and second bands, the cyan ones are of the second and third bands, and the red ones are of the third and fourth bands, as shown in Fig.3(b). Taking the cyan nodal ring as an example, Fig.3(c) shows the phonon dispersion in the $k_x - k_z$ plane near D-L-D′, proving the cyan nodal ring formed by the crossing of the second and third bands. Viewed in the $k_x = 0$ plane, D is the intersecting point of nodal rings, while viewed in the $k_z = \pi$ plane, D is an isolated band touching point that looks like the Dirac point in graphene[47], as shown in Fig.3(d).

To further confirm that the four-fold degenerate states are $|\phi_+\rangle$, $|\phi_-\rangle$, $\mathcal{S}|\phi_+\rangle$, and $\mathcal{S}|\phi_-\rangle$ as discussed in our theory, we examine the eigenvectors of phonons, i.e., the atom vibration directions. Mirror eigenvalue +1 means the state is mapped to itself by the mirror symmetry, so the atoms must vibrate in the $x = 0$ plane. Conversely, mirror eigenvalue -1

corresponds to atoms vibrating out of the $x = 0$ plane. The atom vibration directions of four degenerate bands at the D point are found to be two in-planes and two out-of-planes, verifying our prediction.

Next, we calculate the Berry phase $\gamma$ along a closed path encircling the nodal ring

$$\gamma = i \oint \sum_{n \in occ.} \langle u_{n,k} | \frac{\partial}{\partial k} | u_{n,k} \rangle dk. \tag{4}$$

The summation is done on occupied bands. All nodal rings have a Berry phase of $\pi$, indicating linear dispersion in the neighborhood of nodal rings[31]. Note that when both inversion and time-reversal symmetry is preserved, the Berry phase along any closed path is a multiple of $\pi$, so the integration path can arbitrarily deform as long as the gap is not closed. While in the case of wurtzite GaN where inversion symmetry is broken, the integration path is chosen close enough to the nodal ring to get a nearly-quantized Berry phase.

Nodal rings are accompanied by drumhead surface states on the surface[40]. In the presence of chiral symmetry, drumhead states are limited to the same energy, leading to surface superconductivity[48]. In phonon systems without chiral symmetry, the flatness of drumhead surface states is coincidental[49], depending on the surface roughness. We found drumhead surface states on two distinct surfaces, i.e., the [10(-1)0] surface and the [0001] surface. The phonon local density of states (LDOS) on the [10(-1)0] surface is shown in Fig4(a). The path is chosen as $\overline{A}$-$\overline{L}$-$\overline{A}'$. The projection of the D point corresponds to the cone-shaped LDOS, verifying the semimetal-like point formed by band inversion. Drumhead surface states emanate from $\overline{D}$ in both directions, while those of a single nodal ring only spread in one direction of the nodal ring. The number of surface state branches also multiplies since D belongs to three nodal rings simultaneously. This further increases the surface density of states at $\overline{D}$ and benefits experimental verification. However, the drumhead states are not flat in energy due to the lack of chiral symmetry.

It is worth noting that previously reported nodal links or nodal chains are located on orthogonal mirror-symmetric planes, in both electrons[16] and phonons[50]. Drumhead surface states, emerging with nodal rings, are bounded by the projection of nodal rings on the surface Brillouin zone. However, mirror planes in wurtzite GaN are rotational-symmetric, which enables us to find a surface that is perpendicular to all mirror planes, i.e., the [0001] surface. The projections of mirror planes are one-dimensional lines on the [0001] surface. The projections of nodal rings are discrete line segments. However, thanks to the intersection of nodal rings, segments from different mirror planes (the cyan lines in Fig.3(a)) still form a net on the [0001] surface Brillouin zone. Surface arc states connecting the nodal net on this surface are found, as shown in Fig.4(c). Moreover, the connection is valid in a wide region in $k$ space. We choose two paths, D-D′ and E-E′, shown in Fig.4(b), where D is $(0.384, 0)$ and E is $(0.3, 0)$ in reciprocal lattice units. The phonon surface LDOS along them exhibits robust surface states, as shown in Fig.4(d)-(e).

The shape of the planar nodal chains in wurtzite GaN can be abstracted into multiple intersecting rings, as shown in Fig.5(d), although they are not perfect circles in the realistic phonon spectrum. This simplified description can be generalized to other candidate space groups, where the planar nodal chains lie in the mirror invariant plane, as plotted in Fig.5(a)-(c). Note that space groups 36, 63, and 67 all have the base-centered orthorhombic Bravais lattice, so they share the same first Brillouin zone shape. The same applies to space groups 185, 186, 193, and 194, which share the hexagonal Bravais lattice. In addition, the screw symmetry in space group 67 is $\mathcal{S}_y$, while that in other candidate space groups is $\mathcal{S}_z$, resulting in the difference in the direction of the planar nodal chains, as shown in Fig.5(b) and (c).

## 5. Conclusions

In summary, we first identified a class of symmetry-enforced planar nodal chains in non-symmorphic phononic systems. The symmetry constraints of the planar nodal chain include mirror symmetry $\mathcal{M}_x$, time-reversal symmetry $\mathcal{T}$, and non-symmorphic screw symmetry $\mathcal{S}_{2z}$. We identified 8 out of 230 space groups with proper symmetries to host the planar nodal chain. Using wurtzite GaN as an example, we illustrate the nodal chain phonons in detail based on first-principles calculations. Phonon surface density of states is calculated on the [10(-1)0] and the [0001] surface using iterative Green's function method. The former is inclined to the mirror plane while the latter is perpendicular to the mirror plane. Drumhead surface states are found inside the nodal ring, outside the nodal ring, and connecting different nodal rings. Our study reveals a class of symmetry-enforced planar nodal chain phonons in non-symmorphic materials and expands the catalog of topological nodal chains and related surface states.

# Data Availability

The data that support the findings of this study are available from the corresponding author upon reasonable request.

# Acknowledgments

This work was supported by the National Natural Science Foundation of China (Nos. 51825601 and U20A20301).

# References

1. Bansil, A., Lin, H. & Das, T. Colloquium: Topological band theory. *Rev. Mod. Phys.* **88**, 021004 (2016).

2. Haldane, F. D. M. Nobel Lecture: Topological quantum matter. *Rev. Mod. Phys.* **89**, 040502 (2017).

3. Qi, X.-L. & Zhang, S.-C. Topological insulators and superconductors. *Rev. Mod. Phys.* **83**, 1057–1110 (2011).

4. Armitage, N. P., Mele, E. J. & Vishwanath, A. Weyl and Dirac semimetals in three-dimensional solids. *Rev. Mod. Phys.* **90**, 015001 (2018).

5. Wang, Z. *et al.* Dirac semimetal and topological phase transitions in A3Bi (A=Na, K, Rb). *Phys. Rev. B* **85**, 195320 (2012).

6. Liu, Z. K. *et al.* Discovery of a three-dimensional topological Dirac semimetal, Na3Bi. *Science* **343**, 864–867 (2014).

7. Weng, H., Fang, C., Fang, Z., Bernevig, B. A. & Dai, X. Weyl semimetal phase in noncentrosymmetric transition-metal monophosphides. *Phys. Rev. X* **5**, 011029 (2015).

8. Lv, B. Q., Qian, T. & Ding, H. Experimental perspective on three-dimensional topological semimetals. *Rev. Mod. Phys.* **93**, 025002 (2021).

9. Wan, X., Turner, A. M., Vishwanath, A. & Savrasov, S. Y. Topological semimetal and Fermi-arc surface states in the electronic structure of pyrochlore iridates. *Phys. Rev. B* **83**, 205101 (2011).

10. Young, S. M. *et al.* Dirac semimetal in three dimensions. *Phys. Rev. Lett.* **108**, 140405 (2012).

11. Yang, B.-J. & Nagaosa, N. Classification of stable three-dimensional Dirac semimetals with nontrivial topology. *Nat. Commun.* **5**, 4898 (2014).

12. Gao, Z., Hua, M., Zhang, H. & Zhang, X. Classification of stable Dirac and Weyl semimetals with reflection and rotational symmetry. *Phys. Rev. B* **93**, 205109 (2016).

13. Fang, C., Chen, Y., Kee, H.-Y. & Fu, L. Topological nodal line semimetals with and

without spin-orbital coupling. *Phys. Rev. B* **92**, 081201 (2015).

14. Kim, Y., Wieder, B. J., Kane, C. L. & Rappe, A. M. Dirac line nodes in inversion-symmetric crystals. *Phys. Rev. Lett.* **115**, 036806 (2015).

15. Weng, H. *et al.* Topological node-line semimetal in three-dimensional graphene networks. *Phys. Rev. B* **92**, 045108 (2015).

16. Bzdušek, T., Wu, Q., Rüegg, A., Sigrist, M. & Soluyanov, A. A. Nodal-chain metals. *Nature* **538**, 75–78 (2016).

17. Bi, R., Yan, Z., Lu, L. & Wang, Z. Nodal-knot semimetals. *Phys. Rev. B* **96**, 201305 (2017).

18. Liang, Q.-F., Zhou, J., Yu, R., Wang, Z. & Weng, H. Node-surface and node-line fermions from nonsymmorphic lattice symmetries. *Phys. Rev. B* **93**, 085427 (2016).

19. Wu, W. *et al.* Nodal surface semimetals: Theory and material realization. *Phys. Rev. B* **97**, 115125 (2018).

20. Burkov, A. A., Hook, M. D. & Balents, L. Topological nodal semimetals. *Phys. Rev. B* **84**, 235126 (2011).

21. Liu, Y., Xu, Y., Zhang, S.-C. & Duan, W. Model for topological phononics and phonon diode. *Phys. Rev. B* **96**, 064106 (2017).

22. Süsstrunk, R. & Huber, S. D. Observation of phononic helical edge states in a mechanical topological insulator. *Science* **349**, 47–50 (2015).

23. He, C. *et al.* Acoustic topological insulator and robust one-way sound transport. *Nat. Phys.* **12**, 1124-+ (2016).

24. He, C. *et al.* Acoustic analogues of three-dimensional topological insulators. *Nat. Commun.* **11**, 2318 (2020).

25. Pirie, H., Sadhuka, S., Wang, J., Andrei, R. & Hoffman, J. E. Topological phononic logic. *Phys. Rev. Lett.* **128**, 015501 (2022).

26. Zhang, T. *et al.* Double-Weyl phonons in transition-metal monosilicides. *Phys. Rev. Lett.* **120**, 016401 (2018).

27. Li, J. *et al.* Coexistent three-component and two-component Weyl phonons in TiS, ZrSe, and HfTe. *Phys. Rev. B* **97**, 054305 (2018).

28. Tang, D.-S. & Cao, B.-Y. Topological effects of phonons in GaN and AlGaN: A potential perspective for tuning phonon transport. *J. Appl. Phys.* **129**, 085102 (2021).

29. Zhang, T. T. *et al.* Phononic Helical Nodal Lines with PT Protection in MoB2. *Phys. Rev. Lett.* **123**, 245302 (2019).

30. Li, J. *et al.* Phononic Weyl nodal straight lines in MgB2. *Phys. Rev. B* **101**, 024301 (2020).

31. Liu, G., Jin, Y., Chen, Z. & Xu, H. Symmetry-enforced straight nodal-line phonons. *Phys. Rev. B* **104**, 024304 (2021).

32. Wang, R. Y., Chen, Z. J., Huang, Z. Q., Xia, B. W. & Xu, H. Classification and materials realization of topologically robust nodal ring phonons. *Phys. Rev. Mater.* **5**, 084202 (2021).

33. Wang, X. *et al.* Symmetry-enforced ideal lanternlike phonons in the ternary nitride Li6WN4. *Phys. Rev. B* **104**, L041104 (2021).

34. Wang, J. *et al.* Coexistence of zero-, one-, and two-dimensional degeneracy in tetragonal SnO2 phonons. *Phys. Rev. B* **104**, L041107 (2021).

35. Liu, Q.-B., Wang, Z.-Q. & Fu, H.-H. Ideal topological nodal-surface phonons in RbTeAu-family materials. *Phys. Rev. B* **104**, L041405 (2021).

36. Miao, H. *et al.* Observation of double Weyl phonons in parity-breaking FeSi. *Phys. Rev. Lett.* **121**, 035302 (2018).

37. Chen, Y. S., Huang, F. F., Zhou, P., Ma, Z. S. & Sun, L. Z. Ideal topological phononic nodal chain in K2O materials class. *New J. Phys.* **23**, 103043 (2021).

38. Chen, Z. J., Xie, Z. J., Jin, Y. J., Liu, G. & Xu, H. Hybrid nodal-ring phonons with hourglass dispersion in AgAlO2. *Phys. Rev. Mater.* **6**, 034202 (2022).

39. Zhou, F. *et al.* Hybrid-type nodal ring phonons and coexistence of higher-order quadratic nodal line phonons in an AgZr alloy. *Phys. Rev. B* **104**, 174108 (2021).

40. Zhu, J. *et al.* Symmetry-enforced nodal chain phonons. *Npj Quantum Mater.* **7**, 1–6 (2022).

41. Kresse, G. & Furthmüller, J. Efficient iterative schemes for ab initio total-energy calculations using a plane-wave basis set. *Phys. Rev. B* **54**, 11169–11186 (1996).

42. Kresse, G. & Joubert, D. From ultrasoft pseudopotentials to the projector augmented-wave method. *Phys. Rev. B* **59**, 1758–1775 (1999).

43. Perdew, J. P., Burke, K. & Ernzerhof, M. Generalized gradient approximation made simple. *Phys. Rev. Lett.* **77**, 3865–3868 (1996).

44. Togo, A. & Tanaka, I. First principles phonon calculations in materials science. *Scr. Mater.* **108**, 1–5 (2015).

45. Sancho, M. P. L., Sancho, J. M. L., Sancho, J. M. L. & Rubio, J. Highly convergent schemes for the calculation of bulk and surface Green functions. *J. Phys. F Met. Phys.* **15**, 851–858 (1985).

46. Wu, Q., Zhang, S., Song, H.-F., Troyer, M. & Soluyanov, A. A. WannierTools: An open-source software package for novel topological materials. *Comput. Phys. Commun.* **224**, 405–416 (2018).

47. Castro Neto, A. H., Guinea, F., Peres, N. M. R., Novoselov, K. S. & Geim, A. K. The electronic properties of graphene. *Rev. Mod. Phys.* **81**, 109–162 (2009).

48. Chen, Y., Lu, Y.-M. & Kee, H.-Y. Topological crystalline metal in orthorhombic perovskite iridates. *Nat. Commun.* **6**, 6593 (2015).

49. Liu, Q.-B., Wang, Z.-Q. & Fu, H.-H. Topological phonons in allotropes of carbon. *Mater. Today Phys.* **24**, 100694 (2022).

50. Jin, Y. J. *et al.* Ideal intersecting nodal-ring phonons in bcc C8. *Phys. Rev. B* **98**, 220103 (2018).

Table. 1. The complete list of all space groups that can host planar nodal chains.

| Bravais lattice | Space group number | Space group symbol | Generators | Typical material |
|---|---|---|---|---|
| Orthorhombic | 26 | $Pmc2_1$ | $\mathcal{S}_{2z}, \mathcal{M}_x$ | $CaF_2^*$ |
| | 36 | $Cmc2_1$ | $\mathcal{S}_{2z}, \mathcal{M}_x$ | $Si_2N_2O$ |
| | 63 | $Cmcm$ | $\mathcal{S}_{2z}, \mathcal{M}_x$ | $CaSi^*$ |
| | 67 | $Cmme$ | $\mathcal{S}_{2y}, \mathcal{M}_x$ | $FeSe^*$ |
| Hexagonal | 185 | $P6_3cm$ | $\mathcal{S}_{2z}, \mathcal{M}_x$ | $Li_2PS_3^*$ |
| | 186 | $P6_3mc$ | $\mathcal{S}_{2z}, \mathcal{M}_x$ | $GaN$ |
| | 193 | $P6_3/mcm$ | $\mathcal{S}_{2z}, \mathcal{M}_x$ | $TiCl_3^*$ |
| | 194 | $P6_3/mmc$ | $\mathcal{S}_{2z}, \mathcal{M}_x$ | $NaS$ |

* Unstable materials.

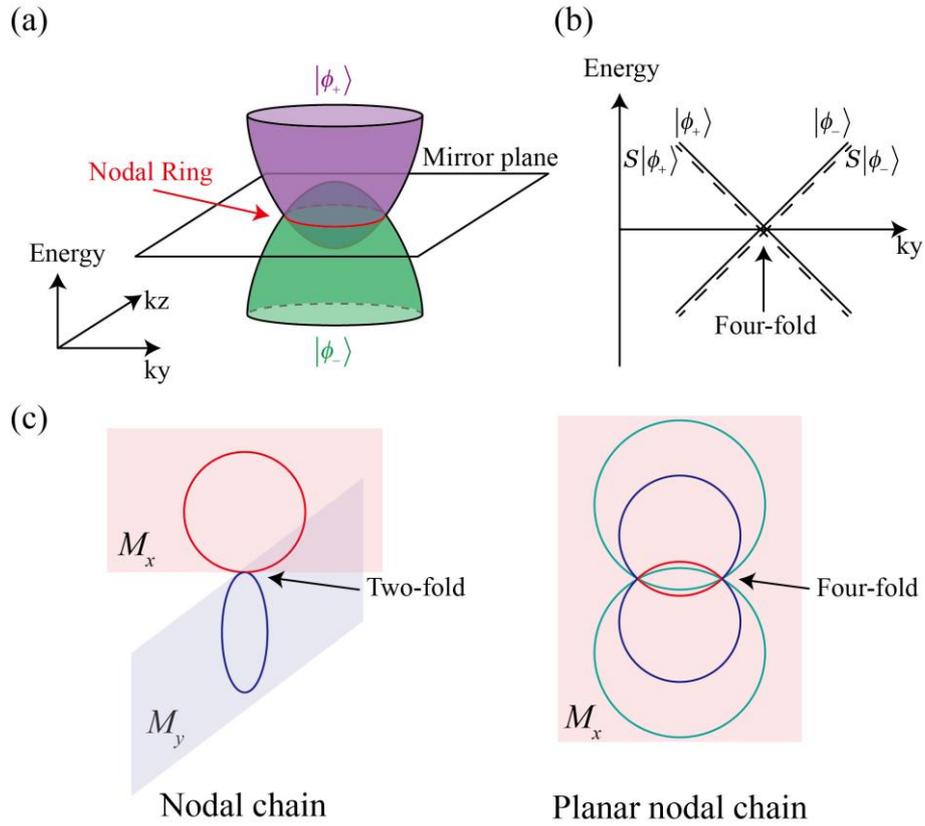

Figure. 1. (a). Sketch of the nodal ring formed on the mirror plane where $k_x = 0$. $|\phi_+\rangle$ and $|\phi_-\rangle$ are forbidden to hybridize. (b). Four orthogonal states $|\phi_+\rangle$, $|\phi_-\rangle$, $S|\phi_+\rangle$, and $S|\phi_-\rangle$ generating a four-fold degenerate point in the $(k_x = 0, k_z = \pi)$ high symmetry line. (c). Illustration of the nodal chain[16,40] and the planar nodal chain (this work). $M_x$ and $M_y$ denote mirror planes. Solid lines denote nodal rings.

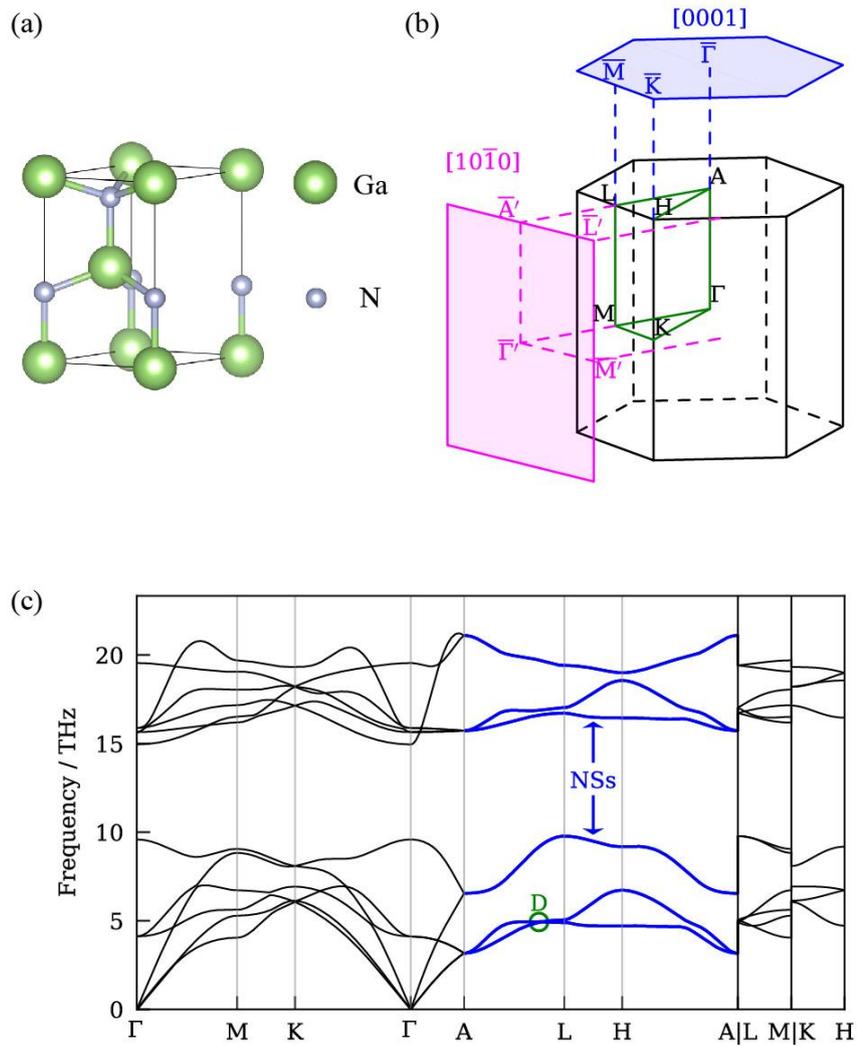

Figure. 2. (a). The lattice structure of wurtzite GaN. (b). First Brillouin zone of bulk GaN and its projection on the [0001] surface (blue) and the [10(-1)0] surface (magenta). (c). Phonon dispersions along high symmetry paths of wurtzite GaN. Nodal surfaces (NSs) degeneracies are denoted by solid blue lines. The green circle marks the crossing point of nodal rings, D.

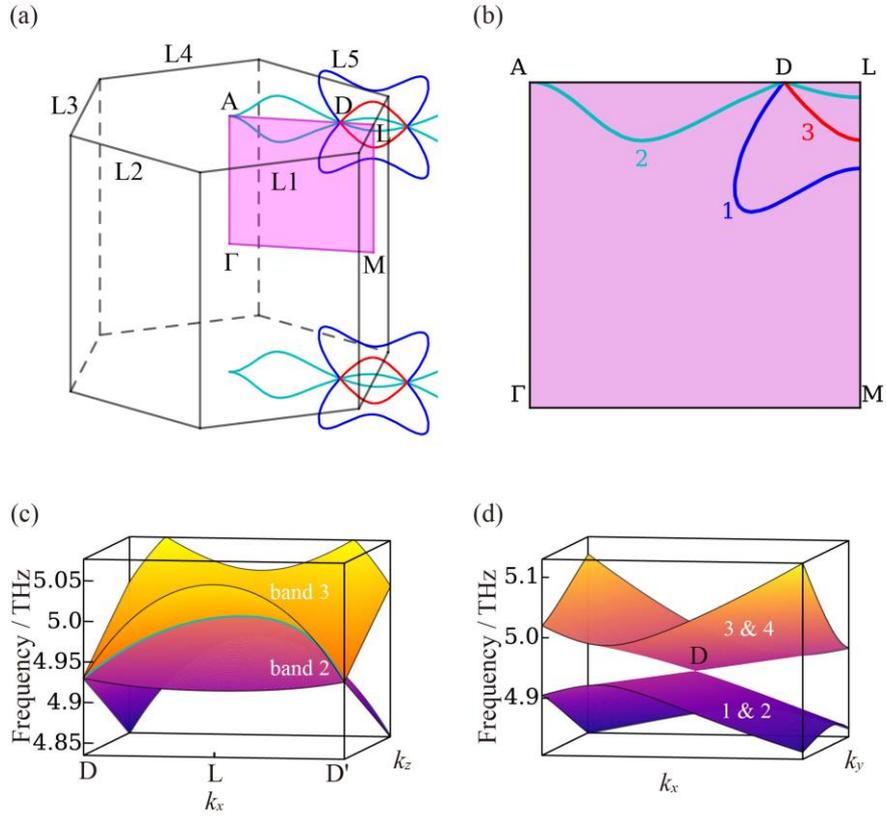

Figure. 3. (a) The planar nodal chains in $k$ space. The red, blue, and cyan lines denote three kinds of nodal rings and D is their crossing point. For clarity, the planar nodal chains in five other directions (A-L1, A-L2, A-L3, A-L4, and A-L5) are not plotted, since they have the same shape due to six-fold screw symmetry. (b) The planar nodal chains plotted in the Γ-A-L-M plane. Numbers denote the occupation numbers of the nodal rings. For example, the cyan line is labeled 2, which means it is the degenerate line of the second and third bands. (c) Phonon dispersion of the second and third band in the $k_x = 0$ plane. The planar nodal chains formed by their crossing are represented by cyan lines. (d) Phonon dispersion near D point in the $k_z = \pi$ plane. The first and second branches are degenerate. The third and fourth branches are degenerate. D is an isolated four-fold degenerate point in this plane.

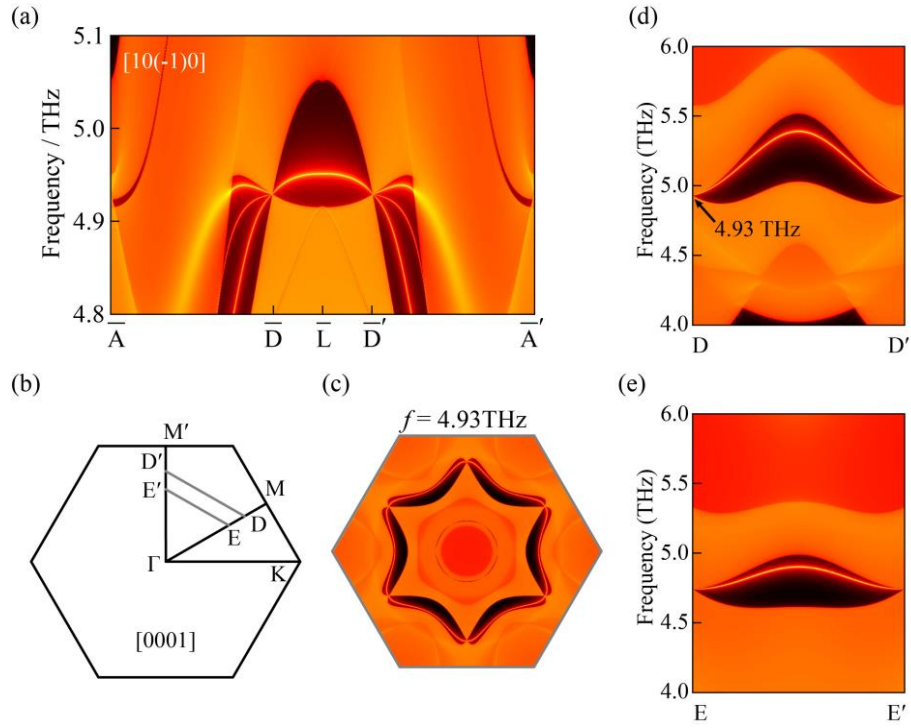

Figure. 4. (a). Phonon local density of states on the [10(-1)0] surface of wurtzite GaN along the $\overline{A}$-$\overline{L}$-$\overline{A}'$ path. Surface states emanating from the surface projection of D points are visible. (b). The [0001] surface Brillouin zone along with two k-paths, D-D′ and E-E′. (c). Iso-frequency surface at 4.93 THz on the [0001] surface Brillouin zone. (d)-(e). Phonon local density of states along D-D′ and E-E′, proving that the surface state is valid in a wide region in $k$ space.

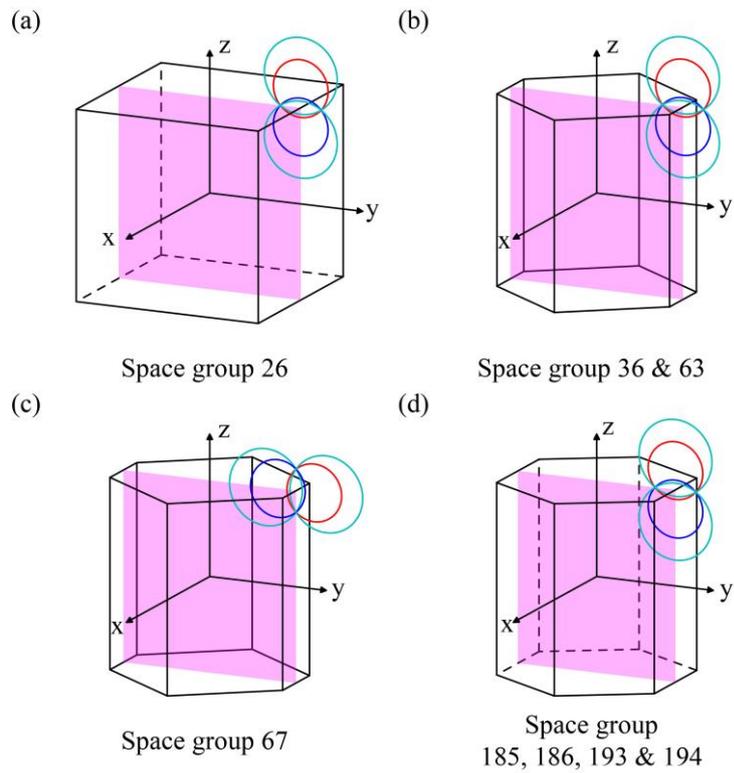

Figure. 5. Schematic diagrams for the planar nodal chains in eight candidate space groups. (a) Space group 26. (b) Space groups 36 and 63. (c) Space group 67. (d) Space groups 185, 186, 193, and 194. The magenta planes denote mirror invariant planes in which the planar nodal chains lie. The red, blue, and cyan circles indicate the nodal rings that build up the planar nodal chains. For clarity, only one situation of the planar nodal chain is plotted in each subfigure.